\documentclass[12pt,a4paper]{article}

\usepackage{color,tikz}
\usepackage[unicode,bookmarks,bookmarksopen,bookmarksopenlevel=2,colorlinks,linkcolor=blue,citecolor=green]{hyperref}

\usepackage{amsmath,eucal,amssymb,amsthm,amsfonts}
\usepackage{mathrsfs,graphicx,texdraw}

\newtheorem{remark}{Remark}

\def\neprod{\setbox0=\hbox{$\nearrow$}
  \box0\kern-1.6em\prod} 
\def\swprod{\setbox0=\hbox{$\swarrow$}%
  \,\,\raise.03em\box0\kern-1.18em\prod} 
\def\seprod{\setbox0=\hbox{$\searrow$}%
  \,\,\raise.00em\box0\kern-2.0em\prod} 

\def\openone{\leavevmode\hbox{\small1\kern-3.3pt\normalsize1}}

\def\re{\mathrm{Re\,}}

\def\bbbe{{\Bbb E}}
\def\bbbc{{\Bbb C}}
\def\bbbr{{\Bbb R}}
\def\bbbz{{\Bbb Z}}

\def\p{{\boldsymbol p}}
\def\q{{\boldsymbol q}}

\newtheorem{proposition}{Proposition}

\begin{document}

\begin{center}
{\LARGE \bf On affine Toda field theories related to  ${\bf D}_r$ \\[6pt] algebras
and their real Hamiltonian forms}

\bigskip

{\bf Vladimir S. Gerdjikov$^{a,b,}$\footnote{E-mail: {\tt gerjikov@inrne.bas.bg}} and
Georgi  G. Grahovski$^{c,}$\footnote{E-mail: {\tt grah@essex.ac.uk}} }

\end{center}

\medskip

\noindent
{\it $^{a}$ Institute of Mathematics and Informatics, Bulgarian Academy of Sciences,
 8 Acad. G. Bonchev str.,  1113 Sofia, Bulgaria }\\[5pt]
{\it $^{b}$Institute for Advanced Physical Studies, 111 Tsarigradsko chaussee,
Sofia 1784, Bulgaria}\\[5pt]
{\it $^{c}$ Department of Mathematical Sciences, University of Essex, Wivenhoe Park, Colchester CO4 3SQ, UK}\\[5pt]

\begin{flushright}
{\it To the memory of our colleague and\\ friend A. B. Yanovski  (1953--2023)}
\end{flushright}

\begin{abstract}
\noindent
The paper deals with affine 2-dimensional Toda field theories related to  simple Lie algebras of the classical series ${\bf D}_r$.   
We demonstrate that the complexification procedure followed by a restriction to a specified real Hamiltonian form
commutes  with the external automorphisms of $\mathfrak{g}$. This is illustrated on the examples
${\bf D}_{r+1}^{(1)} \to {\bf B}_r^{(1)}$ and ${\bf D}_4^{(1)} \to {\bf G}_2^{(1)}$ using external automorphisms of the corresponding extended Dynkin diagrams.

\end{abstract}


\section{Introduction}\label{sec:intro}

It is well known that with each simple Lie algebra $\mathfrak{g}$ of rank $r$ one can relate a 2-dimensional affine Toda field theory (ATFT) \cite{OlPerMikh,BowCor2,BowCor3,Olive,Olove2,AMV,BBT}. Their Lagrangian densities are given by:
\begin{equation}\label{eq: ATFT-Lagr}
{\cal L}[{\bf q}] = {1\over 2}\left( (\partial_{x_0} {\bf q} \cdot \partial^{x_0} {\bf q})-
(\partial_{x_1} {\bf q} \cdot \partial^{x_1} {\bf q} \right) - \sum_{k=0}^{n} n_k \left({\rm e}^{ -2(\alpha_k \cdot{\bf q})} -1\right),
\end{equation}
where the field ${\bf q}(x,t)$ is an $r$-dimensional vector.  The equations of motion are therefore:
\begin{equation}\label{eq:ATFT}
{\partial ^2 {\bf q}  \over \partial x \partial t } = \sum_{j=0}^{r} n_j \alpha_j e^{-2(\alpha_j ,{\bf q}(x,t))},
\end{equation}
where $t= \frac{1}{2}(x_1 -x_0)$ and $x= \frac{1}{2}(x_1 +x_0)$.
The important results that mattered for the derivation of ATFT were: 1) the work of A. Zamolodchikov \cite{Zamol} on deformation of conformal field theories  preserving
integrability and 2) the discovery of Mikhailov's reduction group which allowed one to derive their Lax representation \cite{Mikh,OlPerMikh}.
Here $r$ is the rank of ${\frak g}$, $\alpha_k$'s ($k = 1,\dots,n$) are the simple roots of ${\frak g}$ and $\alpha_0$ is the minimal root.

 The notion of real Hamiltonian forms was introduced in \cite{2} and used to study reductions of ATFTs in \cite{GG,GG2,GGS}.
Real Hamiltonian forms (RHF) are another type of ``reductions'' of Hamiltonian systems. First one complexifies the ATFT and then applies
an involution $\mathcal{C}$ which is compatible with the Poisson structure of the initial system. This is similar to the obtaining a real forms of
a semi-simple Lie algebra. that is why the RHF have indefinite kinetic energy quadratic form.

The structure of the paper is as follows. Section 2 contains some facts about the ATFT and simple Lie algebras.
Section 3 describes the complexification procedure and Section 4 specifies the RHF of ATFT for which the
involution is the external automorphism of $\mathfrak{g}$. Thus we demonstrate that the complexification procedure  commutes with the external
automorphisms of $\mathfrak{g}$.  This is illustrated on the examples ${\bf D}_{r+1}^{(1)} \to {\bf B}_r^{(1)}$ and ${\bf D}_4^{(1)} \to {\bf G}_2^{(1)}$.

\section{Preliminaries}\label{sec:prelim}

The Lax representations of the ATFT models widely discussed in the literature (see e.g. \cite{Mikh,OlPerMikh,Olive,SasKha,SasKha2} and the
references therein). Most of the results on it are related mostly to the normal real form of the Lie
algebra $\mathfrak{ g}$, see \cite{Helg,Bourb1,Cart,Kac,Xu,Xu2,TJ22}. Below we use the following Lax pair:
\begin{eqnarray}\label{eq:2.1}
L\psi \equiv \left(  i{d  \over dx } - iq_x(x,t) - \lambda J_0\right) \psi (x,t,\lambda )=0, \\
M\psi \equiv \left(  i{d  \over dt } -  {1\over \lambda} I(x,t)\right) \psi (x,t,\lambda )=0,
\end{eqnarray}
whose potentials take values in $\mathfrak{g} $. Here $q(x,t) \in \mathfrak{h}$ - the Cartan subalgebra of
$\mathfrak{g}$, and $\q(x,t)=(q_1,\dots , q_r) $ is its dual $r$-component vector. The potentials of the Lax operators take the form
\begin{equation}\label{eq:2.2}
J_0 = \sum_{\alpha \in \pi}^{} E_{\alpha },\qquad I(x,t) =\sum_{\alpha \in \pi}^{} e^{-(\alpha ,\q(x,t))} E_{-\alpha }.
\end{equation}
Here $\pi_{\mathfrak{g}} $ stands for the set of admissible roots of $\mathfrak{g} $, i.e.  $\pi_{\mathfrak{g}} = \{\alpha _0, \alpha
_1,\dots, \alpha _r\} $, with $\alpha _1,\dots, \alpha _r $ being the simple roots of $\mathfrak{ g}$ and $\alpha _0 $ being the
minimal root of $\mathfrak{ g} $.   The Dynkin graph corresponding to the set of
admissible roots $\pi_{\mathfrak{g}}=\{\alpha _0, \alpha _1,\dots,\alpha _r\} $ of ${\frak g}$ is called extended Dynkin diagram (EDD).
The equations of motion are of the form (\ref{eq:ATFT}) where $n_j $ are the minimal positive  integer coefficients $n_k $ that
provide the decomposition of  $\alpha _0 $ over the simple roots of $\mathfrak{g} $:
\begin{equation}\label{eq:n_k}
-\alpha _0 = \sum_{k=1}^{r} n_k\alpha _k.
\end{equation}
It is  well known that ATFT models are an infinite-dimensional Hamiltonian system.  The (canonical)
Hamiltonian structure is given by:
\begin{eqnarray}\label{eq:H-g}
H_{\mathfrak{g}} &=& \int_{-\infty }^{\infty }dx\, \mathcal{H}_{\mathfrak{g}}(x,t), \qquad
\mathcal{H}_{\mathfrak{g}}(x,t)= { 1\over 2} ({\bf p}(x,t),
{\bf p}(x,t))+\sum_{k=0}^{r} n_k(e^{-({\bf q}(x,t),\alpha _k)}-1)  ,\\
\label{eq:ome-g}
\Omega _{\mathfrak{g}} &=& \int_{-\infty }^{\infty } dx\,
\omega_{\mathfrak{g}} (x,t), \qquad
\omega_{\mathfrak{g}} (x,t)=( \delta {\bf p}(x,t)\wedge \delta {\bf q}(x,t)),
\end{eqnarray}
where $H_{\mathfrak{g}} $ is the canonical Hamilton function and $\Omega _{\mathfrak{g}}$ is the canonical symplectic structure.
Here also ${\bf p} = d{\bf q}/d x$ are the canonical momenta
and coordinates satisfying canonical Poisson brackets:
\begin{equation}\label{eq:ATFT-PB}
\{ q_k(x,t) , p_j(y,t)\} = \delta_{jk} \delta (x-y).
\end{equation}
The infinite-dimensional phase space $\mathcal{M}=\{{\bf q}(x,t),{\bf p}(x,t)\}$ is spanned by the canonical coordinates and momenta.

\begin{remark}\label{rem:}
 In what follows we shall need the coefficients $n_k$ for the ${\bf D}_{r+1}^{(1)}$ and $ {\bf B}_r^{(1)}$ algebras, which are equal to
 \cite{Helg,Bourb1,Cart}:
 \begin{equation}\label{eq:}\begin{aligned}
  &{\bf D}_{r+1}^{(1)}: &\qquad  n_1 &=1, \; n_2 = n_3 = \cdots =n_{r-1} =2, &\quad n_r &= n_{r+1}=1, \\
  & {\bf B}_r^{(1)}:    & \qquad n_1 &=1, \; n_2 = n_3 = \cdots =n_{r-1} =n_r =2.
 \end{aligned}\end{equation}
\end{remark}

\section{Complexification of ATFT}\label{ssec:rHF}

The starting point in the construction of real Hamiltonian forms (RHF) is the complexification of the field
functions ${\bf q}(x,t)$ and ${\bf p}(x,t)$ involved in the Hamiltonian \eqref{eq:H-g}:
\[
{\bf q}^\bbbc = {\bf q}^0 + i{\bf q}^1, \qquad {\bf p}^\bbbc =
{\bf p}^0 + i{\bf p}^1.
\]
Next we introduce an involution $\mathcal{ C} $ acting on the
phase space $\mathcal{ M} \equiv \{q_k(x), p_k(x)\}_{k=1}^{n} $ as
follows:
\begin{eqnarray}\label{eq:Cc}
&& \mbox{1)} \qquad \mathcal{ C}(F(p_k,q_k)) = F(\mathcal{ C}(p_k),
\mathcal{ C}(q_k)),  \nonumber\\
&& \mbox{2)} \qquad \mathcal{ C}\left( \{ F(p_k,q_k), G(p_k,q_k)\}\right) =
\left\{ \mathcal{ C}(F), \mathcal{ C}(G) \right\} , \\
&& \mbox{3)} \qquad \mathcal{ C}(H( p_k,q_k)) = H(p_k,q_k) . \nonumber
\end{eqnarray}
Here $F(p_k,q_k) $, $G(p_k,q_k) $ and the Hamiltonian $H(p_k,q_k) $ are
functionals on $\mathcal{M} $ depending analytically on the fields
$q_k(x,t) $ and $p_k(x,t) $.

The complexification of the ATFT is rather straightforward. The resulting
complex ATFT (CATFT) can be written down as standard Hamiltonian system with
twice as many fields ${\bf q}^a(x,t) $, ${\bf p}^a(x,t)  $, $a=0,1 $:
\begin{equation}\label{eq:qp-c}
{\bf p}^\bbbc (x,t) = {\bf p}^0(x,t)+i {\bf p}^1(x,t), \qquad
{\bf q}^\bbbc (x,t)= {\bf q}^0(x,t)+i {\bf q}^1(x,t),
\end{equation}
\begin{equation}\label{eq:qp-pb}
\{{q}_{k}^0(x,t), {p}_{j}^0(y,t) \}= - \{{q}_{k}^1(x,t),
{p}_{j}^1(y,t) \} = \delta _{kj} \delta (x-y).
\end{equation}
The densities of the corresponding Hamiltonian and symplectic form equal
\begin{eqnarray}\label{eq:H_0}
\mathcal{H}_{\rm ATFT}^\bbbc &\equiv & \re \mathcal{H}_{\rm ATFT} ({\bf p}^0+i {\bf p}^1, {\bf q}^0+i {\bf q}^1) \nonumber\\
&=&  {1\over 2 } ({\bf p}^0,{\bf p}^0) -{1\over 2 } ({\bf p}^1,{\bf p}^1) + \sum_{k=1}^{r}n_k e^{-({\bf q}^0,\alpha _k)}
\cos (({\bf q}^1,\alpha _k)) ,  \\
\label{eq:ome_0}
\omega^\bbbc &=& (d{\bf p}^0\wedge  i d{\bf q}^0) - (d{\bf p}^1\wedge
d {\bf q}^1).
\end{eqnarray}
The family of RHF then are obtained from the CATFT by imposing an
invariance condition with respect to the involution
$\tilde{\mathcal{ C}} \equiv \mathcal{ C}\circ \ast $ where by
$\ast $ we denote the complex conjugation. The involution
$\tilde{\mathcal{ C}} $ splits the phase space $\mathcal{ M}^\bbbc
$ into a direct sum $\mathcal{ M}^\bbbc \equiv {\cal M}_+^\bbbc
\oplus \mathcal{M}_-^\bbbc$ where
\begin{equation}\label{eq:M-c}\begin{aligned}
\mathcal{M}_+^\bbbc &= \mathcal{ M}_0 \oplus i \mathcal{ M}_1, &\quad \mathcal{M}_-^\bbbc &= \mathcal{ M}_0 \oplus -i \mathcal{ M}_1, \\
 \mathcal{C}(\q^+ + i \q^-) &= (\q^+ - i \q^-), &\quad  \mathcal{C}(\p^+ + i \p^-) &= (\p^+ - i \p^-).
\end{aligned}\end{equation}
Each involution $\mathcal{C} $ on ${\cal M}$ induces an involution  $\mathcal{C}^\# $ on $\mathfrak{g} $.
Thus to each involution $\mathcal{ C} $ one can relate a RHF of the ATFT.  Due to Property 3), $\mathcal{C}^\# $
preserves the system of admissible roots of $\mathfrak{g} $ (and thus the
extended Dynkin diagrams of $\mathfrak{ g} $).

Indeed, the condition 3) above requires that:
\begin{equation}\label{eq:*1}
(\mathcal{C}(\q ),\alpha ) = (\q , \mathcal{C}^\# (\alpha )), \qquad
\alpha \in \pi_{\mathfrak{g}},
\end{equation}
and therefore we must have $\mathcal{C}(\pi_{\mathfrak{g}}) =
\pi_{\mathfrak{g}} $.

The relation (\ref{eq:*1}) defines uniquely the relation between
$\mathcal{C} $ and $\mathcal{C}^\# $. Using $\mathcal{C}^\#  $ we can
split the root space $\bbbe^n $ into direct sum $\bbbe^n =\bbbe_+\oplus
\bbbe_- $ of two eigensubspaces of $\mathcal{C}^\# $. Taking the average
of the roots $\alpha _j $ with respect to  $\mathcal{C}^\# $ we get:
\begin{equation}\label{eq:*2}
\beta _j = {1  \over 2 } (\alpha _j+\mathcal{C}^\#(\alpha _j)), \qquad
j=0,\dots, n_+=\dim \bbbe_+.
\end{equation}
By construction the set $\{\beta _0,\beta_1,\dots , \beta _{n_+}
\}$ will be a set of admissible roots for some Kac-Moody algebra
with rank $n_+ $. Graphically each set of admissible roots can be
represented by an extended Dynkin diagrams. Therefore one can
relate an automorphism $\mathcal{C}^\# $ to each $\bbbz_2 $
symmetry of the extended Dynkin diagram.

The splitting of $\bbbe^n $ naturally leads to the splittings of the
fields:
\begin{eqnarray}\label{eq:*3}
{\bf p} ={\bf p}^+  +{\bf p}^- , \qquad {\bf q} ={\bf q}^+  +{\bf q}^- ,
\end{eqnarray}
where ${\bf p}^+ ,{\bf q}^+  \in \bbbe_+$ and ${\bf p}^- ,{\bf  q}^-  \in \bbbe_-$. If
we also introduce:
\begin{equation}\label{eq:*4}
\gamma _j = {1  \over  2} (\alpha _j - \mathcal{C}^\#(\alpha _j)), \qquad
j=0,\dots , n_- =\dim \bbbe_-.
\end{equation}
The Hamiltonian along with the terms related to the simple roots, contains also the minimal root $\alpha_0$ given in \eqref{eq:n_k}.
The RHF of ATFT are more general integrable systems than the models
described in \cite{Evans, Evans2, Evans3, SasKha,SasKha2} which involve only the fields ${\bf q}^+ $, ${\bf p}^+
$ invariant with respect to $\mathcal{C} $.

\section{RHF of ATFT from external automorphisms}\label{ssec:Dr-1}

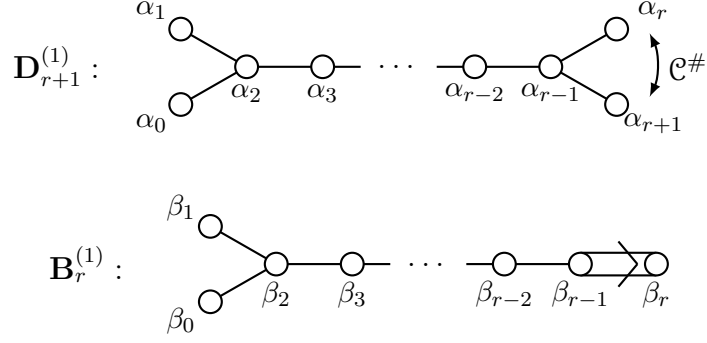
\begin{figure}
\begin{center}
\begin{tikzpicture}
\draw[thick] (-0.866,0.5) --(0,0);
\draw[thick] (-0.866,-0.5) --(0,0);
\draw[thick] (0,0) --(1.5,0);
\draw[thick] (2.5,0) --(4,0);
\draw[thick] (4.866,0.5) --(4,0);
\draw[thick] (4.866,-0.5) --(4,0);


\draw[thick,fill=white] (-0.866,0.5) circle (1.5 mm);
\draw[thick,fill=white] (-0.866,-0.5) circle (1.5 mm);
\draw[thick,fill=white] (0,0) circle (1.5 mm);
\draw[thick,fill=white] (1.0,0) circle (1.5 mm);
\draw (2,0) node {$\dots$};

\draw[thick,fill=white] (3,0) circle (1.5 mm);
\draw[thick,fill=white] (4.0,0) circle (1.5 mm);
\draw[thick,fill=white] (4.866,0.5) circle (1.5 mm);
\draw[thick,fill=white] (4.866,-0.5) circle (1.5 mm);

\draw (-1.25,0.75) node {\small $\alpha_{1}$};
\draw (-1.25,-0.75) node {\small $\alpha_{0}$};
\draw (0,-0.35) node {\small $\alpha_2$};
\draw (1.0,-0.35) node {\small $\alpha_3$};
\draw (3,-0.35) node {\small $\alpha_{r-2}$};
\draw (4.0,-0.35) node {\small $\alpha_{r-1}$};
\draw (5.35,0.75) node {\small $\alpha_{r}$};
\draw (5.35,-0.75) node {\small $\alpha_{r+1}$};

\draw[thick,latex-latex] (5.3,-0.4) arc (-25:25:1.0);

\draw (-2.5,0) node {${\bf D}_{r+1}^{(1)}:$};
\draw (5.8,0) node {${\cal C}^\#$};

\end{tikzpicture}

\vspace{0.2in}

\begin{tikzpicture}
\draw[thick] (-0.866,0.5) --(0,0);
\draw[thick] (-0.866,-0.5) --(0,0);
\draw[thick] (0,0) --(1.5,0);
\draw[thick] (2.5,0) --(4,0);
\draw[thick] (4,0.15) --(5,0.15);
\draw[thick] (5,-0.15) --(4,-0.15);


\draw[thick,fill=white] (-0.866,0.5) circle (1.5 mm);
\draw[thick,fill=white] (-0.866,-0.5) circle (1.5 mm);
\draw[thick,fill=white] (0,0) circle (1.5 mm);
\draw[thick,fill=white] (1.0,0) circle (1.5 mm);
\draw (2,0) node {$\dots$};

\draw[thick,fill=white] (3,0) circle (1.5 mm);
\draw[thick,fill=white] (4.0,0) circle (1.5 mm);
\draw[thick,fill=white] (5,0) circle (1.5 mm);

\draw (-1.25,0.75) node {\small $\beta_{1}$};
\draw (-1.25,-0.75) node {\small $\beta_{0}$};
\draw (0,-0.4) node {\small $\beta_2$};
\draw (1.0,-0.4) node {\small $\beta_3$};
\draw (3,-0.4) node {\small $\beta_{r-2}$};
\draw (4.0,-0.4) node {\small $\beta_{r-1}$};
\draw (5.0,-0.4) node {\small $\beta_{r}$};

\draw[thick] (4.5,-0.3) --(4.75,0);
\draw[thick] (4.5,0.3) --(4.75,0);

\draw (-2.5,0) node {${\bf B}_r^{(1)}:$};

\end{tikzpicture}
\end{center}
  \caption{\small Extended Dynkin diagrams of the  complex untwisted affine Kac-Moody algebras ${\bf D}_{r+1}^{(1)}$ and ${\bf B}_r^{(1)}$ (upper and lower panels respectively).}\label{fig:Dynkin-DB}
\end{figure}

Let us now outline the RHF obtained from the algebras ${\bf D}_{r+1}$ using the external automorphisms.
The set of admissible roots ${\bf D}_{r+1}$ are:
\begin{eqnarray*}\label{eq:Dr-roots}
\alpha_1&=& e_1-e_2, \quad \alpha_2=e_2-e_3, \quad \dots, \quad \alpha_{r}=e_{r}-e_{r+1}, \quad
\alpha_{r+1}=e_{r}+e_{r+1}, \quad \alpha_0=e_1+e_2.
\end{eqnarray*}
Here $\alpha_1, \alpha_2\dots , \alpha_{r+1}$ form the set of simple roots of
${\bf D}_{r+1}^{(1)}$ and $\alpha_0$ is the minimal root of the algebra. The extended Dynkin diagram of ${\bf D}_{r+1}^{(1)}$ is shown on the upper panel of Figure \ref{fig:Dynkin-DB}.
Here also $e_k$, $k=1,\dots,r+1$ are the basic vectors of the dual Euclidean space ${\Bbb E}^{r+1}$, i.e.
any vector ${\bf q} \in {\Bbb E}^{r+1}$ can be written as
\[
{\bf q}=\sum_{k=1}^{r+1}q_ke_k.
\]
The components $q_k$, $k=1,\dots r+1$ are the coordinates in the phase space.

The fundamental weights of ${\bf D}_r^{(1)}$ are
\begin{eqnarray*}
\omega_k&=& e_1+\dots +e_k, \qquad 1\leq k\leq r-1,\\
\omega_{r}&=& {1\over 2}(e_1+e_2+\dots + e_{r-1}+ e_{r}-e_{r+1}), \\
 \omega_{r+1}&=& {1\over 2}(e_1+e_2+\dots + e_{r-1}+ e_{r}+e_{r+1}). \nonumber
\end{eqnarray*}
The corresponding Hamiltonian has the form
\begin{equation}\label{eq:Ham-Dr}
H_{{\bf D}_r^{(1)}}=\int_{-\infty}^{\infty} {\rm d}x\, \left(\sum_{k=1}^{r}{p_k^2\over 2}+ \sum_{k=1}^{r} n_k {\rm e}^{2( q_k-q_{k-1})}
+{\rm e}^{2(q_{r+1}-q_r)} +{\rm e}^{-2(q_{r+1}+q_r)} + {\rm e}^{2(q_1+q_2)}\right)
\end{equation}

\medskip
\subsection{From ${\bf D}_{r+1}^{(1)}$ to ${\bf B}_{r}^{(1)}$}
\noindent
Let us  choose  ${\frak g}\simeq {\bf D}_{r+1}^{(1)}$ and fix up the
involution $\mathcal{C} $ acting on the phase space as follows:
\begin{eqnarray*}\label{eq:ex4}
\mathcal{C} (q_k) =q_{k}, \quad \mathcal{C} (p_k) =p_{k},\quad k=1,\dots , r; \quad \mathcal{C} (q_{r+1})=-q_{r+1}, \quad
\mathcal{C} (p_{r+1}) =-p_{r+1}.
\end{eqnarray*}
Then introducing on $\mathcal{M}_\pm$ new coordinates by
\begin{eqnarray*}\label{eq:p-pm4}
&& q_k^+=q_k, \qquad  p_k^+=p_k, \qquad q_r^-=q_r, \qquad  p_r^-=p_r,
\qquad k=1,...,r;\\
&& q_{r+1}^{-} =q_{r+1}, \qquad p_{r+1}^{-} =p_{r+1},
\end{eqnarray*}
i.e. $\dim \mathcal{M}_+=2r $ and  $\dim \mathcal{M}_-=2$.

This involution induces an involution $\mathcal{C}^\# $ of the Kac-Moody
algebra ${\bf D}_{r+1}^{(1)} $ which acts on the root space as follows
(see the upper panel of Fig. \ref{fig:Dynkin-DB}):
\begin{eqnarray*}\label{eq:E4.2}
\mathcal{C}^\# (e_k) =e_{k}, \qquad k=1,\dots, r; \qquad
\mathcal{C}^\# (e_{r+1}) =-e_{r+1},\\
\mathcal{C}^\#(\alpha _{k}) =\alpha _{k}, \qquad
\mathcal{C}^\# (\alpha _{r+1}) =\alpha _r, \qquad
\mathcal{C}^\# (\alpha _{r}) =\alpha _{r+1},
\end{eqnarray*}
The involution $\mathcal{C}^\# $ splits the root space
$\bbbe^{r+1} $ into a direct sum of  eigensubspaces:
$\bbbe^{r+1}=\bbbe_+\oplus \bbbe_- $ with $\dim \bbbe_+=r $ and
$\dim \bbbe_-=1 $. The restriction of $\pi $ onto $\bbbe_+ $ leads
to the admissible root system $\pi'=\{ \beta _0,\dots, \beta _r\}
$ of ${\bf B}_{r}^{(1)} $:
\begin{equation*}\label{eq:bet-4}
\beta _k=\alpha _k, \qquad k=0,\dots , r-1; \qquad
\beta _r={1 \over 2} \left[\alpha _r+\mathcal{C}^\#(\alpha _r)\right], \qquad
\end{equation*}
The subspace $\bbbe_- $ is spanned by the only nontrivial vector
\begin{equation*}\label{eq:gam-4}
\gamma _r={1 \over 2} \left[\alpha _r-\mathcal{C}^\#(\alpha _r )\right] = -e_{r+1}.
\end{equation*}
The reduced RHF is described by the  densities
$\mathcal{H}^\bbbr$, $\omega ^\bbbr$:
\begin{eqnarray}
&& \mathcal{H}^\bbbr  = {1\over 2} \sum_{k=1}^{r} p_{k}^{+}{}^2 - {1\over 2}  p_{r+1}^{-}{}^2 +\sum_{k=1}^{r}
n_k e^{2(q_{k+1}^+ -q_k^+)} +  e^{2q_{r+1}^+} \cos q_{r+1}^- , \label{eq:H4H}\\
&& \omega^\bbbr = \sum_{k=1}^{r} \delta p_k^+\wedge \delta
q_k^+ - \delta p_{r+1}^-\wedge \delta q_{r+1}^-.\label{eq:H4om}
\end{eqnarray}
The restriction on ${\Bbb E}_+$ (i.e. setting $q_{r+1}^-=0$ in \eqref{eq:H4H}) gives the canonical Hamiltonian of ${\bf B}_r^{(1)}$ ATFT:
\begin{equation}\label{eq:Ham-Br}
H_{{\bf B}_r^{(1)}}=\int_{-\infty}^{\infty} {\rm d}x\, \left(\sum_{k=1}^{r}{p_k^2\over 2}+ \sum_{k=1}^{r-1} n_k {\rm e}^{2( q_k-q_{k-1})} + 2{\rm e}^{-2q_r} + {\rm e}^{2(q_1+q_2)}\right),
\end{equation}
as expected.

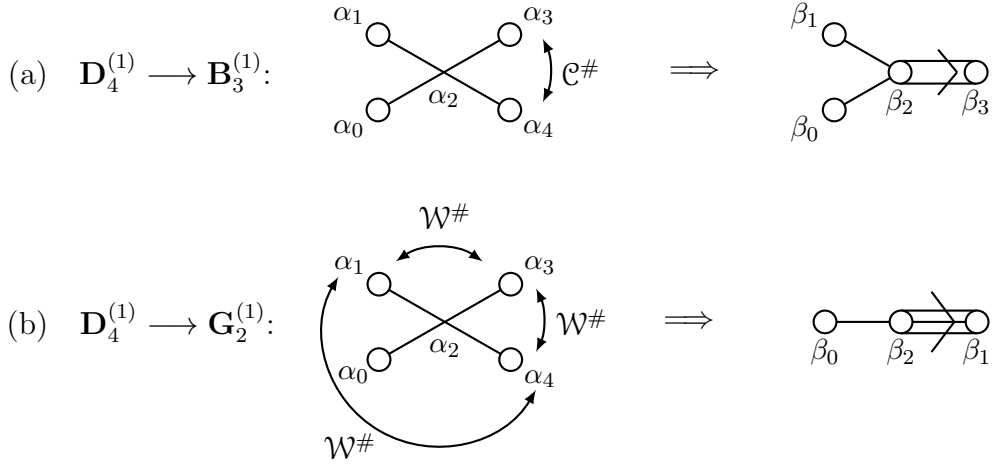
\begin{figure}
\begin{center}
\begin{tikzpicture}
\draw[thick] (-0.866,0.5) --(0,0);
\draw[thick] (-0.866,-0.5) --(0,0);
\draw[thick] (0.866,0.5) --(0,0);
\draw[thick] (0.866,-0.5) --(0,0);


\draw[thick,fill=white] (-0.866,0.5) circle (1.5 mm);
\draw[thick,fill=white] (-0.866,-0.5) circle (1.5 mm);
\draw[thick,fill=white] (0.866,0.5) circle (1.5 mm);
\draw[thick,fill=white] (0.866,-0.5) circle (1.5 mm);

\draw (-1.25,0.75) node {\small $\alpha_{1}$};
\draw (-1.25,-0.75) node {\small $\alpha_{0}$};
\draw (0,-0.35) node {\small $\alpha_2$};
\draw (1.25,0.75) node {\small $\alpha_3$};
\draw (1.25,-0.75) node {\small $\alpha_4$};

\draw[thick,latex-latex] (1.3,-0.4) arc (-25:25:1.0);

\draw (-4,0) node { (a)$ \quad {\bf D}_4^{(1)} \longrightarrow {\bf B}_3^{(1)} $:};
\draw (1.8,0) node {${\cal C}^\#$};

\draw (3.3,0) node {$\Longrightarrow$};

\draw[thick] (5.134,0.5) --(6,0);
\draw[thick] (5.134,-0.5) --(6,0);
\draw[thick] (6,0.15) --(7.0,0.15);
\draw[thick] (6,-0.15) --(7.0,-0.15);


\draw[thick,fill=white] (5.134,0.5) circle (1.5 mm);
\draw[thick,fill=white] (5.134,-0.5) circle (1.5 mm);
\draw[thick,fill=white] (6,0) circle (1.5 mm);
\draw[thick,fill=white] (7.0,0) circle (1.5 mm);

\draw (4.75,0.75) node {\small $\beta_{1}$};
\draw (4.75,-0.75) node {\small $\beta_{0}$};
\draw (6,-0.4) node {\small $\beta_2$};
\draw (7.0,-0.4) node {\small $\beta_3$};

\draw[thick] (6.5,-0.3) --(6.75,0);
\draw[thick] (6.5,0.3) --(6.75,0);

\end{tikzpicture}

\vspace{0.2in}

\begin{tikzpicture}
\draw[thick] (-0.866,0.5) --(0,0);
\draw[thick] (-0.866,-0.5) --(0,0);
\draw[thick] (0.866,0.5) --(0,0);
\draw[thick] (0.866,-0.5) --(0,0);


\draw[thick,fill=white] (-0.866,0.5) circle (1.5 mm);
\draw[thick,fill=white] (-0.866,-0.5) circle (1.5 mm);
\draw[thick,fill=white] (0.866,0.5) circle (1.5 mm);
\draw[thick,fill=white] (0.866,-0.5) circle (1.5 mm);

\draw (-1.25,0.75) node {\small $\alpha_{1}$};
\draw (-1.2,-0.7) node {\small $\alpha_{0}$};
\draw (0,-0.35) node {\small $\alpha_2$};
\draw (1.25,0.75) node {\small $\alpha_3$};
\draw (1.25,-0.75) node {\small $\alpha_4$};

\draw[thick,latex-latex] (1.2,-0.4) arc (-25:25:1.0);
\draw[thick,latex-latex] (0.5,0.8) arc (55:125:1.0);
\draw[thick,latex-latex] (-1.4,0.6) arc (150:330:1.5);

\draw (-4,0) node { (b)$ \quad {\bf D}_4^{(1)} \longrightarrow {\bf G}_2^{(1)} $:};
\draw (1.8,0) node {${\cal W}^\#$};
\draw (0,1.4) node {${\cal W}^\#$};
\draw (-1.25,-1.65) node {${\cal W}^\#$};

\draw (2.8,0) node {$\qquad \Longrightarrow$};

\draw[thick] (5,0) --(6,0);
\draw[thick] (6,0.15) --(6.0,0.15);
\draw[thick] (6,-0.15) --(7.0,-0.15);
\draw[thick] (6,0.15) --(7.0,0.15);
\draw[thick] (6,0) --(7.0,0);

\draw[thick,fill=white] (5,0) circle (1.5 mm);
\draw[thick,fill=white] (6,0) circle (1.5 mm);
\draw[thick,fill=white] (7,0) circle (1.5 mm);

\draw (5,-0.4) node {\small $\beta_{0}$};
\draw (6,-0.4) node {\small $\beta_{2}$};
\draw (7,-0.4) node {\small $\beta_{1}$};

\draw[thick] (6.4,-0.4) --(6.7,0);
\draw[thick] (6.4,0.4) --(6.7,0);

\end{tikzpicture}
\end{center}
  \caption{\small Reductions of $ {\bf D}_4^{(1)}$ affine Lie algebra: {\bf (a)} $ {\bf D}_4^{(1)}\rightarrow  {\bf B}_3^{(1)}$; {\bf (b)} $ {\bf D}_4^{(1)}\rightarrow  {\bf G}_2^{(1)}$.}\label{fig:Dynkin-BG}
\end{figure}

\medskip

\subsection{From ${\bf D}_{4}^{(1)}$ to ${\bf B}_{3}^{(1)}$}

If we take now a particular case od Example 1 with $r=3$, the Hamiltonian \eqref{eq:Ham-Dr} will reduce to the generic ATFT Hamiltonian related to ${\bf D}_4^{(1)}$:
\begin{equation}\label{eq:Ham-D4}
H_{{\bf D}_4^{(1)}}=\int_{-\infty}^{\infty} {\rm d}x\, \left(\sum_{k=0}^{3}{p_k^2\over 2}+ + {\rm e}^{2( q_2-q_{1})} + 2{\rm e}^{2( q_3-q_{2})}
+ {\rm e}^{2( q_4-q_{3})} +{\rm e}^{-2(q_{3}+q_4)} + {\rm e}^{2(q_1+q_2)}\right).
\end{equation}
The corresponding extended root system is
\[
\overline{\pi} \left({\bf D}_4^{(1)}\right)=\{e_1+e_2,e_1-e_2, e_2-e_3, e_3-e_4, e_3+e_4\},
\]
and the extended Dynkin diagram is given on Figure \ref{fig:Dynkin-BG}(a). This algebra has 4 outer ${\Bbb Z}_2$ automorphisms and one ${\Bbb Z}_2$ automorphism \cite{Bourb1,VinOni}. Here we will take $C^\#$ to be the outer ${\Bbb Z}_2$-automorphism, swapping $\alpha_3$ and $\alpha 4$ (See Figure \ref{fig:Dynkin-BG}(a)). The action of $C^\#$ on the root space ${\Bbb E}^4$ is then given by:
\begin{eqnarray*}\label{eq:E4.2a}
\mathcal{C}^\# (e_1) =e_{1},  \qquad
\mathcal{C}^\# (e_2) =e_{2},\qquad \mathcal{C}^\# (e_3) =e_{4}, \qquad \mathcal{C}^\# (e_4) =e_{3};\\
\mathcal{C}^\#(\alpha _{1}) =\alpha _{1}, \qquad \mathcal{C}^\#(\alpha _{2}) =\alpha _{2}, \qquad \mathcal{C}^\#(\alpha _{3}) =\alpha _{4}, \qquad \mathcal{C}^\#(\alpha _{4}) =\alpha _{3}.
\end{eqnarray*}
The involution $\mathcal{C}^\# $ splits the root space
$\bbbe^{r+1} $ into a direct sum of  eigensubspaces:
$\bbbe^{r+1}=\bbbe_+\oplus \bbbe_- $ with $\dim \bbbe_+=3 $ and
$\dim \bbbe_-=1 $. The restriction of $\pi $ onto $\bbbe_+ $ leads
to the admissible root system $\pi'=\{ \beta _0,\beta_1,\beta_2, \beta _3\}
$ of ${\bf B}_{3}^{(1)} $:
\begin{equation*}\label{eq:bet-4db}
\beta _0=\alpha _0, \qquad \beta _1=\alpha _1,  \qquad \beta _2=\alpha _2,
\qquad \beta _3={1 \over 2} \left[\alpha _3+\mathcal{C}^\#(\alpha _3)\right]= e_3.
\end{equation*}
The subspace $\bbbe_- $ is spanned by the vector
\begin{equation*}\label{eq:gam-4db}
\gamma _3={1 \over 2} \left[\alpha _3-\mathcal{C}^\#(\alpha _3)\right] = -e_{4}.
\end{equation*}
As a result,   the  densities of the reduced RHF become:
\begin{eqnarray}
&& \mathcal{H}^\bbbr  = {1\over 2} \left(p_{1}^{+}\right)^2 + {1\over 2} \left(p_{2}^{+}\right)^2 + {1\over 2} \left(p_{3}^{+}\right)^2 - {1\over 2} \left(p_{4}^{-}\right)^2 +{\rm e}^{2(q_{2}^+ -q_1^+)} + 2{\rm e}^{2(q_{3}^+ -q_2^+)} +  2{\rm e}^{-2q_{3}^+} \cos q_{4}^- , \label{eq:H2H}\\
&& \omega^\bbbr = \sum_{k=1}^{r} \delta p_k^+\wedge \delta
q_k^+ - \delta p_{r+1}^-\wedge \delta q_{r+1}^-.\label{eq:H2om}
\end{eqnarray}
The restriction on ${\Bbb E}_+$ (i.e. setting $p_4^-=q_{4}^-=0$ in \eqref{eq:H2H}) gives the canonical Hamiltonian of ${\bf B}_3^{(1)}$ ATFT:
\begin{equation}\label{eq:Ham-B3}
H_{{\bf B}_3^{(1)}}=\int_{-\infty}^{\infty} {\rm d}x\, \left({1\over 2} \sum_{j=1}^{3} \left( p_{j}^{+}\right)^2  +{\rm e}^{2(q_{2}^+ -q_1^+)} + 2{\rm e}^{2(q_{3}^+ -q_2^+)}
 +  2{\rm e}^{-2q_{3}^+}\right).
\end{equation}

\medskip

\subsection{From ${\bf D}_4^{(1)}$ to ${\bf G}_2^{(1)}$}

 If we take again the ${\bf D}_4^{(1)}$ Hamiltonian \eqref{eq:Ham-D4} and use the ${\Bbb Z}_3$ outer automorphism of ${\bf D}_4^{(1)}$ sketched on Figure \ref{fig:Dynkin-BG}(b), then we will get the induced action ${\cal W}^\#$ on the root space ${\Bbb E}^4$:
\begin{equation}\label{eq:a}\begin{split}
{\cal W}^\# (e_1) &= \frac{1}{2} (e_1 +e_2 + e_3 - e_4), \quad {\cal W}^\# (e_2) = \frac{1}{2} (e_1 +e_2 - e_3 + e_4), \\
{\cal W}^\# (e_3) &= \frac{1}{2} (e_1 -e_2 + e_3 + e_4), \quad {\cal W}^\# (e_4) = \frac{1}{2} (e_1 - e_2 - e_3 - e_4),
\end{split}\end{equation}
i.e.
\begin{equation}\label{eq:b}\begin{split}
{\cal W}^\# (e_1 +e_2) = e_1 +e_2, \qquad {\cal W}^\# (e_2 -e_3) = e_2 -e_3, \qquad {\cal W}^\# (e_1 -e_2) = e_3 -e_4, \\
{\cal W}^\# (e_3 -e_4) = e_3 +e_4, \qquad {\cal W}^\# (e_3 +e_4) = e_1 -e_2,
\end{split}\end{equation}
and
\begin{eqnarray*}
\beta_0&=& -(e_1 + e_2), \qquad \beta_2=(e_2 - e_3), \\
 \beta_1 &=&{1\over 3}\left[\alpha_1+ {\cal W}^\#(\alpha_1) +{\cal W}^\#{}^2(\alpha_1)\right]={1\over 3}(e_1-e_2+ 2e_3).
\end{eqnarray*}
This is the extended root system of ${\bf G}_2^{(1)}$ (see Figure \ref{fig:Dynkin-BG}(b)). This results to a RHF related to ${\bf G}_2^{(1)}$ with Hamiltonian given by
\begin{equation*}\label{eq:Ham-G2}
H_{{\bf G}_2^{(1)}}=\int_{-\infty}^{\infty} {\rm d}x\, \left({1\over 2} \left(p_{1}^{+}\right)^2 + {1\over 2} \left(p_{2}^{+}\right)^2 + {1\over 2} \left(p_{3}^{+}\right)^2+ {\rm e}^{-2(\beta_0,{\bf q})} + 2{\rm e}^{-2(\beta_1,{\bf q})} + 3{\rm e}^{-2(\beta_2, {\bf q})}\right),
\end{equation*}
where the three component vectors ${\bf q}$ and ${\bf p}$ are restricted by $q_1+q_2+q_3=0$ and $p_1+p_2+p_3 =0$.

The above results establish the following
\begin{proposition}
Let us consider the RHF of ATFT using the external automorphisms of the algebras ${\bf D}^{(1)}_{r+1} $.
If we use the second order external automorphisms, then
\begin{equation}\label{eq:rhf}\begin{split}
\mathcal{H}^{\Bbb R}_{{\bf D}^{(1)}_{r+1}} = \mathcal{H}_{{\bf B}^{(1)}_r}, \qquad \mathcal{H}^{\Bbb R}_{{\bf D}^{(1)}_{4}} = \mathcal{H}_{{\bf B}^{(1)}_3}.
\end{split}\end{equation}
If we use the third order external automorphism of $D_4$, then we obtain
\begin{equation}\label{eq:c}\begin{split}
\mathcal{H}^{\Bbb R}_{{\bf D}^{(1)}_{4}} = \mathcal{H}_{{\bf G}^{(1)}_2}.
\end{split}\end{equation}

\end{proposition}

The next step would be to analyze the more general situation when the RHF is generated by $\mathcal{C}_0 \circ \mathcal{C}^\#$
where $\mathcal{C}_0$ is a generic suitable automorphism and $\mathcal{C}^\#$ is an external automorphism.
Our hypothesis is that the above Proposition will hold true also for this more general case.


\section{Conclusions}

We presented here real Hamiltonian forms of affine Toda field theories related to the untwisted complex Kac-Moody algebra ${\bf D}_4^{(1)}$. We outlined
the construction of the RHF  and studied  ${\Bbb Z}_2$ and ${\Bbb Z}_3$ symmetries of the
extended Dynkin diagrams. Thus resulted in reductions to ${\bf B}_3^{(1)}$ and ${\bf G}_2^{(1)}$.

The spectral properties of the Lax operators of the real Hamiltonian forms of ATFTs can be studied in the frame of the ISM \cite{FaTa,ZMNP}. This will lead to the construction of Jost solutions and scattering data for Lax operators with complex-valued Cartan elements \cite{BeCo1,BeCo2,GeYa}. The continuous spectrum of the Lax operators will consist of $2h$  rays intersecting at the origin and closing angles $\pi /h$.

The interpretation of the ISM as a generalized Fourier transforms \cite{AKNS,IP2,GeYa} allows one to study all the
fundamental properties of the corresponding  nonlinear evolutionary equations (NLEE's): i)~the description of the class of NLEE related to a~given Lax operator
$L(\lambda)$ and solvable by the ISM; ii)~derivation of the infinite family of integrals of motion;  iii)~their hierarchy
of Hamiltonian structures \cite{GVYa}; and iv) description  of the gauge equivalent systems \cite{GGMV,GGM2,Grah,GraCo,Grah2,Grah3}.

Some additional problems are natural extensions to the results presented here:

\begin{itemize}
  \item The complete classification of all nonequivalent RHF of ATFT.

\item The description of the hierarchy of Hamiltonian structures of RHF of ATFT (for a review of the infinite-dimensional cases see e.g.
\cite{DriSok,67} and the references therein) and the classical $r$-matrix. It is also an
open problem to construct the RHF for ATFT using some of its
higher Hamiltonian structures.

\item The extension of the dressing Zakharov-Shabat method \cite{ZaSha2}  to the above classes of Lax operators is also an open problem. One of the difficulties is due to the fact that the $\mathbb{Z}_h$ reductions requires dressing factors with $2h$ pole singularities  \cite{VSG-88,VG-13}.

\item Another open problem is to study types of boundary conditions and boundary effects of ATFT's and their RHF \cite{Caudrelier,Doikou1}.

\end{itemize}

The last and more challenging problem is to prove the complete integrability of all these models. The ideas of \cite{AKNS, GVYa} about the interpretation of the inverse scattering method as a generalized Fourier transform holds true also for the $\mathbb{Z}_h$ reduces Lax operators \cite{GeYa,SIAM*14,VG-Ya-13,GYa*14}. This may allow one to derive the action-angle variables for these classes of NLEE.

\section*{Acknowledgements}

One of us (V.G.) is grateful to professor P. Bamidis for useful discussions and for his hospitality at Aristotle University, Thessaloniki, Greece, for common
participation under the Operational Programme "Science and Education for Smart Growth", project   BG05M2OP001-2.016-0025.
This work is  supported by the Bulgarian National Science Fund, grant KP-06N42-2.

\end{document}